\documentclass[runningheads,a4paper]{llncs}

\usepackage{amssymb}
\usepackage{graphicx}
\usepackage{epstopdf}
\usepackage{url}
\newcommand{\keywords}[1]{\par\addvspace\baselineskip
\noindent\keywordname\enspace\ignorespaces#1}

\begin{document}

\mainmatter  

\title{Anticipatory Mobile Digital Health: \\Towards Personalised Proactive Therapies\\ and Prevention Strategies}

\titlerunning{Anticipatory Mobile Digital Health}

%
%
\author{Veljko Pejovic\and Abhinav Mehrotra\and Mirco Musolesi}
\authorrunning{Anticipatory Mobile Digital Health}

\author{Veljko Pejovic\inst{1} \and Abhinav Mehrotra\inst{2,3} \and
Mirco Musolesi\inst{3}}
%
%
\tocauthor{Veljko Pejovic (University of Ljubljana, Slovenia),
Abhinav Mehrotra (University of Birmingham, United Kingdom),
Mirco Musolesi (University College London, United Kingdom)}
\institute{University of Ljubljana, Slovenia\\
\and
University of Birmingham, United Kingdom\\
\and
University College London, United Kingdom}

%
%
\toctitle{Lecture Notes in Computer Science}
\tocauthor{Authors' Instructions}
\maketitle

\begin{abstract}

The last two centuries saw groundbreaking advances in the field of healthcare: from the invention of the vaccine to organ transplant, and eradication of numerous deadly diseases. Yet, these breakthroughs have only illuminated the role that individual traits and behaviours play in the health state of a person. Continuous patient monitoring and individually-tailored therapies can help in early detection and efficient tackling of health issues. However, even the most developed nations cannot afford proactive personalised healthcare at scale. Mobile computing devices, nowadays equipped with an array of sensors, high-performance computing power, and carried by their owners at all time, promise to revolutionise modern healthcare. These devices can enable continuous patient monitoring, and, with the help of machine learning, can build predictive models of patient's health and behaviour. Finally, through their close integration with a user's lifestyle mobiles can be used to deliver personalised proactive therapies. In this article, we develop the concept of anticipatory mobile-based healthcare -- \textit{anticipatory mobile digital health} -- and examine the opportunities and challenges associated with its practical realisation.

\end{abstract}

\section{Introduction}

Mobile computing devices, such as smartphones and wearables\footnote{In this paper by \textit{wearables} we refer to smartwatches, smartglasses, e-garments and similar clothing and accessory items equipped with computing and sensing capabilities.} represent more than occasionally used tools, and nowadays coexist with their users throughout the day. In addition, these devices host an array of sensors, such as a GPS receiver, accelerometer, heart rate sensors, microphones and cameras, to name a few~\cite{Lane2010}. When data from these sensors are processed through machine learning algorithms, they can reveal the context in which a device is. The context can include anything from a device's location to a user's physical activity, even stress levels and emotions~\cite{Rachuri2010,Lu2012}. Therefore, the personalisation and the sensing capabilities of today's mobiles can provide a close view of a user's behaviour and wellbeing. 

Above all, mobile devices are always connected. They represent the most direct point of contact for the majority of the world's population. Mobile phones, for example, provide an opportunity for an intimate, timely communication unimaginable just twenty years ago. One of the consequences is that mobile devices are becoming a new channel for the delivery of health and wellbeing therapies. For instance, digital behaviour change interventions (dBCIs) harness smartphones to deliver personally tailored coaching to participants seeking behavioural change pertaining to smoking cessation, depression or weight loss~\cite{Lathia2013}. Communication through a widely used, yet highly personal device ensures that a person can be contacted at all times, which might be crucial in case of suicide prevention interventions. In addition, the smartphone is used for numerous purposes, which protects a dBCI participant from stigmatisation that may happen if the device is used exclusively for therapeutic purposes.

Besides the inference of the current state of the sensed context, an ever-increasing amount of sensor data, advances in machine learning algorithms, and powerful computing hardware packed in mobile devices, allow the predictions of the future state of the context. \textit{Context predictions} have already been shown in the domains of human mobility~\cite{Ashbrook2003_gpsprediction,Scellato2011,Chon2013}, but also population health state~\cite{Madan2010}. Every next generation of mobile devices comes equipped with new sensors, and soon we may expect galvanic skin response (GSR), heart rate, body temperature oxymetry sensors as standard features\footnote{See for example the proposal by Intel:~\url{http://iq.intel.co.uk/glimpse-of-the-future-the-healthcare-smartwatch/}.}. This would open up the ability to accurately predict the health state of an individual.

\textit{Anticipatory mobile computing} is a novel concept that, just like context prediction, relies on mobile sensors to provide information upon which the models of context evolution are built, yet it extends the idea with reasoning and actioning upon such predictions. The concept is inspired by biological systems that often use the past, present and the predicted future state of itself and its environment to change the state at an instant, so to steer the future state in a preferred direction~\cite{Rosen1985}. Anticipatory mobile computing has a potential to revolutionise proactive healthcare. Health and wellbeing problems could be predicted from personalised sensor readings, and preventive actions could be taken even before the onset of a problem. We term this new paradigm -- \textit{Anticipatory Mobile Digital Health}, and in this paper we discuss the challenges and opportunities related to its practical realisation. 
First, we examine the key enablers (i.e., mobile and wearable sensors) that provide the contextual data which can be leveraged to infer the health state of a user. 
Then, we discuss machine learning techniques used for building predictive models of the user's (health) context. We are particularly interested in the models that describe how the context might change after an intervention or a therapy. We investigate the challenges related to unobtrusive learning of the impact of an intervention to a person, and the opportunities for highly personalised healthcare. We also take into account individual differences among users, and the potential for capturing and including genetic pre-determinants into the system. We continue with the examination of human-computer interaction issues related to the therapy delivery, and conclude with a consideration of ethical issues in anticipatory mobile digital health. Finally, while we have examined the potential for inducing a change in a person's behaviour through anticipatory mobile computing before~\cite{Pejovic2014b}, this paper extends the idea on the much larger domain of digital healthcare, and elaborates on particular challenges and opportunities in the area.

\section{Mobile Sensing for Healthcare}
The use of wireless and wearable sensors represents a novel and a rapidly evolving paradigm in healthcare. These sensors have the potential to revolutionise the way of assessing the health of a person. Sensor embedded devices are given to the patients in order to obtain their health related data remotely. These devices do not only help a patient in reducing the number of visits to the clinic, but also offer unprecedented opportunities to the practitioners for diagnosing diseases and tailoring treatments through continuous real-time sampling of their patients' heath data. Furthermore, some of these devices empower the users with the ability to self-monitor and curb certain well-being issues on their own.  

Today's mobile phones are laden with sensors that are able monitor context various modalities such as physical movement, sound intensity, environment temperature and humidity, to name a few. Some previous studies have showed the potential of mobile phones in providing data that can be used to infer the health state of a user~\cite{Consolvo2006Houston,Consolvo2008,Lane2011b,Canzian2015MoodTraces}. Houston~\cite{Consolvo2006Houston} and UbiFit~\cite{Consolvo2008} are the early examples of mobile sensing systems designed to encourage users to increase their physical activity. Houston monitors a user's physical movement by counting the number of steps taken via an accelerometer that serves as a pedometer. Whereas, UbiFit relies on the Mobile Sensing Platform (MSP)~\cite{Lester2006Practical} to monitor varied physical activities of a user. MSP is capable of inferring physical activities including walking, running, cycling, cardio and strength exercise, and other non-exercise physical activities, such as housework. BeWell is a mobile application that continuously monitors a user's physical activity, social interaction and sleeping patterns, and helps the user manage their wellbeing~\cite{Lane2011b}. Bewell relies on sensors such as accelerometer, microphone and GPS, which are embedded in mobile phones. 
%
In~\cite{Canzian2015MoodTraces} the authors show that the depressive states of users can be inferred purely from location and mobility data collected via mobile phones. The above examples demonstrate the close bond between smartphone sensed data and different aspects of human health and well-being.

A particularly interesting example of mobile healthcare monitoring is given by \textit{LifeWatch \footnote{\url{www.lifewatch.com}}}, a smartphone that is equipped with health sensors that constantly monitor the user's vital parameters including ECG, body temperature, blood pressure, blood glucose, heart rate, oxygen saturation, body fat percentage and stress levels. A user has to perform a specific action in order to take health measurements. For example, a user should hold the phone's thermometer against the forehead in order to measure the body temperature, and to take ECG readings, the user should clutch the phone horizontally with a thumb and forefinger placed directly on top of a set of sensors that are placed on the sides of the phone. The sensor data are sent to the cloud for the analysis and the results are delivered back to the user within a short time interval. Such a phone can prove to be extremely useful in the healthcare domain. However, there is still no proof of the accuracy of its results. 

Although within their owners' reach for most of the time, smartphones do not stay in a constant contact physical with the users, and consequently are limited with respect to personal data they can provide. More recently, mobile phone companies have introduced smartwatches that link with mobile phones and enable the users to perform actions on the mobiles without actually interacting with them. These devices open up new possibilities for health data sensing. First, they maintain continuous physical contact with their users, and second, they host a new set of sensors, usually unavailable on traditional smartphones. In general these devices come with the accelerometer, heart rate and body temperature sensors. Smartwatches are inspired by the concept of a smart-wristband, a device that monitors the health state of a user and presents it in a visual form on the linked mobile phone. Smart-wristbands enable real-time health state monitoring, and have achieved a considerable commercial success among health-aware population (e.g. Jawbone\footnote{\url{jawbone.com}}). Initially these bands were able to report only a user's physical activity. However, new sensors, such as body temperature and hearth rate, have been introduced, together with a more sophisticated data analytics and presentation to the user. 

Mobile sensing on the phone is for the majority of readings limited by the amount of physical contact the user makes with the phone. Smartwatches and smart-wristbands ensure that the contact is there, yet are limited to a particular part of the users body -- her wrist. Sensor embedded smart-wearables designed to dedicatedly monitor specific health related parameter from a specific part of a user's body, have appeared recently and promise more reliable sensing. Such smart-wearables could enable healthcare practitioners to obtain their patients' health data continuously and in the natural environment of the patient. These devices come with a variety of health sensors. Pulse and oxygen in blood sensor, airflow sensor, body temperature sensor, electrocardiogram sensor, glucometer sensor, galvanic skin response sensor (GSR), blood pressure sensor (sphygmomanometer), and electromyography sensor (EMG), are some examples of the health sensors embedded in the smart-wearables.  Some examples of smart-wearables include \textit{Epoc Emotiv}~\cite{Fraga2013Emotiv}, an EEG headset capable of capturing brain signals that can be analysed to infer a user's thoughts, feelings, and emotions. \textit{MyoLink} is another wearable that can continuously monitor the user's muscles and heart. It can capture muscle energy output, which in turn can be used to quantify the user's fatigue, endurance and recovery level. Also, it can be placed on the chest to continuously track the heart rate of the user. \textit{ViSi mobile}\footnote{\url{www.visimobile.com}}, worn on a wrist, measures blood pressure, haemoglobin level, heart rate, respiration rate, and skin temperature. The device is highly portable and enables the user to monitor their health at anytime and anywhere. 

The next step in wearable computing is the one in which devices become completely stealth, and as in the Weiser's vision of pervasive computing, completely integrated with people's lives~\cite{Weiser1991}. Shrinking the size of smart-wearables is push in that direction, for example reducing the size of a device from something obtrusive to a small adaptive device that the user can wear on their bodies and forget about it. BioStamp~\cite{Perry2015CheckEngine} is a device composed of small and flexible electronic circuits that stick directly to the skin like a temporary tattoo and monitors the user's health. It is a stretchable sensor capable of measuring body temperature, monitoring exposure to ultraviolet light, and checking pulse and blood-oxygen levels. The company envisions future versions of BioStamp able to monitor changes in blood pressure, analyse sweat, and obtain signals from the user's brain and heart in order to use them in electroencephalograms and electrocardiograms~\cite{Perry2015CheckEngine}. 

These wearable sensors enable the continuous measurement of health metrics and deliver treatment to the patients on time. Yet, the difficulty of continuous monitoring is not the only problem in modern healthcare. Recent studies have shown that around 50\% of the prescribed drugs are never taken~\cite{Nasseh2012Adherence,Osterberg2005Adherence}, and thus, prescribed therapies fail to improve the health of the patients~\cite{Rasmussen2007Adherence}. In order to address this problem, Hafezi et al.~\cite{Hafezi2015Ingestible} proposed \textit{Helius}, a novel sensor for detecting the ingestion of a pharmaceutical tablet or a capsule. The system is basically an integrated-circuit micro-sensor developed for daily ingestion by patients, and as such allows real-time measurement of medication ingestion and adherence patterns. Moreover, Helius enables practitioners to measure the correlation between drug ingestion and patients health parameters, e.g. physical activity, heart rate, sleep quality, and blood pressure, all of which can be  sensed by mobile sensors. 

The ecosystem of devices supporting health sensing is already substantial and constantly increasing. Soon, healthcare practitioners will have a remote multifaceted view of a patient's health in real time. The key enabler is the unobtrusiveness of these smart sensing devices. Furthermore, issues such as the accuracy of measurements, accountability for mistakes and the security of a user's privacy, need to be thoroughly addressed before these devices can penetrate into the official medical practice. 

In this paper we discuss the novel concept of anticipatory mobile digital health, outlining the challenges and opportunities in this promising field. Although smart health sensing devices are still in their infancy, we believe that we will witness a rapid evolution of this research area in the coming years.


\section{Anticipatory Mobile Computing}

Anticipation, for living systems, is the ability to reason upon past, present and predicted future information. Such, a systems Rosen described as \textit{``a system containing a predictive model of itself and/or its environment, which allows it to change state at an instant in accord with the model's predictions pertaining to a later instant"}~\cite{Rosen1985}, thus indicated that there is an internal predictive model that an anticipatory system builds and maintains. The concept of an anticipatory computing system envisions a digital implementation of such a model, and automated actioning based on the model's predictions. Yet, an anticipatory computing system is of interest only if the anticipation carries a value for the end-user.

\begin{figure}
  \centering
  \includegraphics[width=\linewidth]{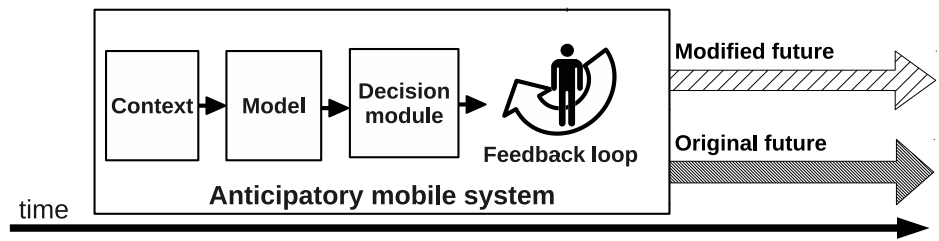}
  \caption{Anticipatory mobile systems predict context evolution and the impact that current actions can have on the predicted context. The feedback loop consisting of a mobile and a human enables the system to affect the future.}
  \label{fig:feedback_loop}
\end{figure}

We argue that modern mobile computing devices fulfil the necessary prerequisites for anticipatory computing. First, thanks to built-in sensors and personalised usage these devices can gather the information about theirs, and indirectly the user's state, and the state of the environment; second, their computing capabilities allow devices to build predictive models of the evolution of the state; finally, the bond between a device and its end-user is so tight that automated suggestions (based on the anticipation) a device might convey to a user, are likely to influence the user's actioning. After all, people already look into their smartphones when they need to navigate in a new environment or choose a restaurant. To clarify the concept of anticipation on mobile devices (termed \textit{Anticipatory Mobile Computing}), in Figure~\ref{fig:feedback_loop} we sketch a system that senses the context and builds a model of the environment evolution, which gives it the original predicted future. The system then evaluates the possible outcome of its actions on the future. An action that leads to the preferred modified future is realised through the feedback loop that involves interaction of the system with the user.

\section{Anticipatory Healthcare System Architecture Design}

\begin{figure}
  \centering
  \includegraphics[width=0.8\linewidth]{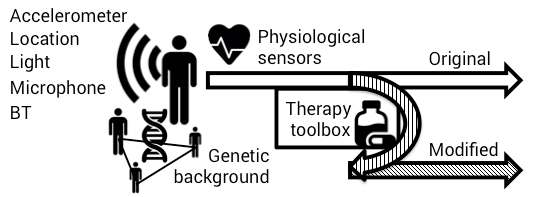}
  \caption{Anticipatory mobile systems predict context evolution and the impact their actions can have on the predicted context. The feedback loop consisting of a mobile and a human enables the system to affect the future.}
  \label{fig:therapy_feedback_loop}
\end{figure}

The opportunity to infer the health and well-being state of an individual with the help of mobile sensing, together with the perspective of anticipatory mobile computing, pave the way for preventive healthcare through anticipatory mobile healthcare systems. We sketch the main ideas behind such a system in Figure~\ref{fig:therapy_feedback_loop}. Physiological (e.g. heart rate, GSR) and conventional mobile sensors (e.g. GPS, accelerometer) provide training data for machine learning models of the context (e.g. a user's depression level) and its evolution. The models predict the future state of the context, termed the original future, and the state after an intervention or a therapy, termed the modified future. Based on the predictions, a therapy with the most preferred outcome is selected and conveyed to the user. Finally, different users may react differently to the same therapy, and close sensor-based patient monitoring, together with a-priori inputs, such as a user's genetic background, are used to custom tailor the therapies.

A practical realisation of an anticipatory mobile digital health system requires that the following building blocks are present:
\begin{itemize}
\item \textbf{Mobile sensing.} The role of this block is to manage which of a number of available mobile sensors are sampled, and how often. Mobile devices' sensors were originally envisioned as occasionally used features, and their frequent sampling can quickly deplete a device's battery. At the same time, important events may be missed if sampling is too coarse. 
\item \textbf{Therapy and prevention toolbox.} This block contains definitions of possible therapies and prevention strategies that can be delivered to the user. Although in future we envision further automatisation of this module, for now, we feel that a professional therapist's expertise should be harnessed to limit the number of possible therapies, and oversee their deployment.
\item \textbf{Machine learning core.} Anticipatory mobile digital health employs machine learning for two separate aspects of health state evolution modelling: \textit{context evolution model} and \textit{therapy/prevention-effect model}. The former connects sensor data with higher-level context, and provides a predictive model of how the context might evolve. The latter provides a picture of how different therapies might affect a user's health state. We discuss these models in detail in the next section.
\item \textbf{User interaction interface.} The success of an anticipatory mobile digital healthcare system is limited by the user's compliance with the provided therapy and prevention strategy. The look, feel and the behaviour of the mobile application that delivers the therapy or prevention strategy to a user is crucial in this step. in the following section, we also discuss the challenges in designing a successful user interaction interface.
\end{itemize}




\section{Challenges and Opportunities}


Numerous challenges obstruct the path towards implementations of anticipatory mobile digital healthcare systems. Rooted in mobile sensing, anticipatory mobile digital health faces challenges such as resource, primarily energy, inefficiency of continuous sensing, and the difficulty of reliable context modelling. Yet, these challenges are common for a larger field of mobile sensing, and a thorough discussion on these issues is available elsewhere~\cite{Lane2010,Pejovic2015,Klasnja2012}. Instead, here we focus on aspects that are unique to anticipatory mobile healthcare. The use of machine learning algorithms to model and predict user behaviour and the effect of a therapy or a prevention strategy on the future health state of a specific user is the main challenge. The value of machine learning models, for instance, increases with the amount of available training data for her. 
Second, the mobile monitors the user, and may suggest therapies, yet, it is the user herself that decides whether to take the therapy or to follow certain preventive measures or not. Besides machine learning, future anticipatory mobile digital health developers should pay a special attention to the human-computer interaction issues in this field, and try to answer -- what is the best way to convey an advice/therapy to a user, so that the compliance with the proposed therapy or prevention strategy is the highest? Finally, the area of ethics, responsibility and entity roles in anticipatory mobile digital health remains an uncharted territory. In the rest of the section we discuss each of the challenges individually, and provide positional guidelines for overcoming the challenges.

\subsection{Machine Learning in Anticipatory Mobile Digital Health}
Anticipatory mobile digital health, as stated in the previous section, employs machine learning for two separate aspects of health state evolution modelling: \textit{context evolution model} and \textit{therapy/prevention-effect model}. First, a model of a user current and predicted future health state is needed. In this model, a relationship between mobile sensor data and high-level health state is built. The model can be direct, if certain values of physiological sensor readings indicate a certain health state, or indirect, if sensor readings reveal contextual aspects that can be connected to a health state of an individual -- for instance, GPS readings can reveal user mobility, which in turn can hint a user's depression state~\cite{Canzian2015MoodTraces}. In the next step, inference models are extended to provide predictions of the future state of the health state, either directly, or indirectly through the predictions of the future context. Forecasting user's next location is an active area of research, with substantial achievements~\cite{Ashbrook2003_gpsprediction,Scellato2011,Chon2013}. For many other aspects of the user's behaviour and health state, reliable predictive models still do not exist, and even the possibility of them being built remains an open question.

The second major machine learning model in anticipatory mobile digital health is the model of the impact of a possible therapy or prevention strategy on the predicted future health state of a user. There are two non-exclusive ways to construct such a model: one is to harness the existing expertise in healthcare to map available therapies to health state transitions. For example, we could map antidepressants to a transition from depressive states to a healthy state. However, these rules are not suitable for preventive healthcare. Anticipatory mobile digital health operates on predictions, and consequently therapies should aim to \textit{prevention}. In addition, although mobile devices remain highly personal, and the sensor data uncovers fine-grained individual health state information, these general rules limit the ability of the system to deliver personalised healthcare. An alternative approach is to build a therapy/prevention-effect model by monitoring the evolution of a user state after a proactive therapy or prevention strategy is delivered. By comparing the original predicted state with the actual state recorded some time after the therapy (or prevention strategy), we can identify the relationship between the therapy (or prevention strategy) and the future health state change. Built this way, a model reveals successful proactive therapies, which is difficult to achieve in the traditional practice. Moreover, what works for one patient may not work for another -- these models are highly personalised, and can reveal therapies that are useful for a particular kind of a person only. Still, we argue that these models should not be built from the scratch -- the available therapies that could be automatically suggested to a particular patient in a particular situation should be determined by the rules stemming from the existing medical expertise.

\textbf{Learning with a user.}
Automated tool-effect modelling in anticipatory mobile digital health requires that a therapy (or prevention strategy) is induced to a user so that its effects can be observed. This outcome is then used to train and refine the model. \textit{Reinforcement learning} where an agent uses a tool in the intervention environment (which for example can be represented through a Markov decision process) is a natural way to model the problem~\cite{Sutton1998}. In every step, a certain tool is selected, used, and the observed change in the health state elicits a reward that reflects how positive the change is. 

\textbf{Measuring health state.}
Thus, there is a need for a suitable metric for \textit{measuring the health state change}. Here we need to evaluate the effect of a proactive therapy or prevention strategy, basically compare the \textit{original predicted health state} and the \textit{modified predicted health state}. We argue that the comparison metric has to be domain dependent. For example, if an anti-stress therapy is evaluated, the difference between the predicted skin conductivity and heart rate values without an intervention, and the actual values after the intervention, is a reasonable measure of stress level change~\cite{Healey2005}. However, system designers should have in mind that the metric has to be both suitable for machine learning algorithms as well as relevant from the healthcare point of view.

\textbf{Learning without interfering.}
Reinforcement learning uncovers the mapping between therapies and health state changes. Delivering a previously unused therapy or prevention strategy refines the model, as we learn more about how the user reacts to this tool. From the practical point of view, however, we face a dilemma: use a tool that is known to result in a positive health change outcome, or experiment with an unused tool that might yield an even better outcome. In reinforcement learning this dilemma is known as \textit{exploration vs. exploitation trade-off}. Strategies for solving the dilemma in an anticipatory mobile digital health setting should be aware of the possible irreversible negative consequences of a wrong therapy or prevention strategy. Preferably, the system should learn as much as possible without explicit delivery of therapies to a user. Such a learning concept is called \textit{latent learning}. It is a form of learning where a subject is immersed into an unknown environment or a situation without any rewards or punishments associated to them~\cite{Tavris1997}. Latent learning has been demonstrated in living beings who form a cognitive map of the environment solely because they are immersed into the environment, and later use the same map in decision making. We argue that mobile computing devices, through multimodal sensing, can harness latent learning to build a model of the user reaction with respect to certain actions or environmental changes that correspond to ones targeted by the therapies. This is particularly relevant for therapies that are not based on medications, such as behavioural change interventions~\cite{Pejovic2014b}. For example, suppose a depression prevention system can provide the user with the suggestion to go out for a dinner with friends. We can get an a priori knowledge of how this suggestion would affect the user, for example if on a separate occasion we detect that the participant went out for a dinner with friends, and we gauge the depression levels, estimated through mobility and physical activity metrics, before and after the dinner. Defining how the expected action -- going out with friends -- should manifest from the point of view of sensors -- e.g., a number of Bluetooth contacts detected, location, time of the day -- is one of the prerequisites for practical latent learning. Again, interdisciplinary efforts are crucial to ensure that the detected reaction corresponds to the one that should be elicited by the tool. 

\subsection{Personalised Healthcare}

Current therapies are often created as ``one size fits all", yet in many cases individuals react differently. For example, antidepressants are ineffective in 38\% of the population, while cancer drugs work for only one quarter of the patient population~\cite{PersonalizedMedicine2014}. Personalised therapies promise to revolutionise healthcare, by avoiding the traditional trial-and-error therapy prescription, minimising adverse drug reactions, revealing additional or alternative uses for medicines and drug candidates~\cite{Mancinelli2000}, and curbing the overall cost of healthcare~\cite{PersonalizedMedicine2014}.  

Anticipatory mobile digital health is poised to bring personalised healthcare closer to mainstream practice. Not only can mobile sensing provide a glimpse into individual behavioural patterns, identifying risky lifestyles, but therapy/prevention-effect machine learning models can also take into account a patient's genetics in order to individualise the therapy or prevention strategy. Investigation of which genes impact the occurrence and reaction to a treatment of a certain disease is a very active area of research. The potential for healthcare improvement is immense, having in mind that with some conditions, such as melanoma tumors, the majority of cases are driven by certain person-specific genetic mutations, and could be targeted by specific drugs~\cite{PersonalizedMedicine2014}.
The relationship is not one way, and anticipatory mobile digital health could also help with pharmacogenomics, the study of how genes affect a person's response to drugs. Identifying common pieces of genetic background in populations who reacted to an anticipatory therapy or prevention strategy in the same way would help find the relationship between genes and health treatments. Finally, the inclusion of the genetic background in the common medical practice is not far from reality -- in 2014 a human genome sequencing for less than USD \$1000 became available.


\subsection{HCI Issues in Anticipatory Mobile Digital Health}

Despite the automation that anticipatory mobile digital health brings, in the end, it is up to a user to comply with the given therapy or prevention strategy. This is particularly important for behavioural change intervention therapies, that are delivered in cases where the health state is directly influenced by patient's behaviour. Consequently, the communication between the system and the patient has to be seamless. Users are an important part of the system, and their inclusion requires an appropriate interface between the participant and the system. As noted by Russell et al.~\cite{Russell2003}, a system that autonomously brings decisions and evolves over the course of its lifetime needs to be transparent to the user. Through the user interface such a system must be understandable by the user and capable of review, revision, and alteration. In addition, the content should be framed to emphasise that the tool can help, yet it is fundamental to avoid to harass and patronise the participant.

The timing of a therapy or a prevention strategy is also important for its successful delivery. This is particularly true for automated therapies delivered via a mobile device. An inappropriately timed intervention that comes, for instance, when a patient is in a meeting, or riding a bicycle, may lead to annoyance, or may be completely overlooked by the patient. Mobile sensing helps with identifying opportune moments to deliver therapies. The context in which a user is, such as her location, physical activity and engagement in a task, to an extent determines her interruptibility~\cite{Pejovic2014,Mehrotra2015NotifyMe}. Machine learning and mobile sensing is harnessed for monitoring a user's reaction to an interruption arriving when the user is in a certain context, and from there on a model of personal interruptibility is built. Querying the model with a momentarily value of a user's context returns the estimated interruptibility at the moment. While practical implementations of the above models already exist~\cite{Pejovic2014}, in future we envision predictive models of user interruptibility. Finally, we highlight that opportune moments denote those time at which a patient is likely to quickly acknowledge/read the content of a delivered message. Identifying moments at which the delivered information will have the highest medical impact is even more important, yet due to the difficulty of getting the training data (we would need to deliver the same therapy or prevention strategy at different times to the same user) identifying such moments remains very challenging.

\subsection{Ethics and Accountability}

Privacy issues in mobile sensing emerged soon after the proliferation of smartphones started about a decade ago. Misuse and leaking of information that can be collected by a mobile device, such as a user's location, collocation with other people, physical activity of a user, may deter people from trusting mobile application. Trust is a key component for the success of anticipatory mobile digital health applications, and every care should be taken that personal information does not leak. Ensuring that sensor data do not leave the device at which they were collected is one way to minimise the risk. However, this complicates the  construction of joint machine learning models discussed earlier.

The responsibility chain in the domain of anticipatory mobile digital health is yet to be defined. Unsuccessful therapies can have serious consequences. It is unclear who is to blame if a delivered therapy or prevention strategy does not improve the health state of a patient, or even worse, endangers the person's life. A therapist who designed the therapy, a software architect who devised the underlying machine learning components, and the patient herself, all play a role in the process.

\section{Conclusions}

Personalised and proactive healthcare brings undisputed benefits in terms of therapy (or prevention strategy) efficiency and cost effectiveness of the healthcare system. Mobile devices have a potential to become both our most vigilant observers, and closest advisors. Anticipatory mobile digital health harnesses the sensing capabilities of mobiles to learn about the user health state and predict its evolution, so that proactive therapies tackling predicted health issues are deliver to the user in advance. With the help of machine learning that takes into account rich sensor data and a user's genetic background, anticipatory mobile digital health applications can tailor personalised therapies. Yet, in addition, the concept can be used to learn more about how therapies affect different demographics, users who behave in a certain way, or have a particular genetic background. Generalising from a larger pool of users and therapies can identify groups for which a therapy (or prevention strategy) is successful, basically uncovering new facts about drugs. Finally, we believe anticipatory mobile health applications warrant a discussion on their inclusion into the health insurance frameworks. 

\bigskip

{\bf \noindent Acknowledgements}

This work was supported by the EPSRC grants ``UBhave: ubiquitous and social computing for positive behaviour change" (EP/I032673/1) and ``Trajectories of Depression: Investigating the Correlation between Human Mobility Patterns and Mental Health Problems by means of Smartphones'' (EP/L006340/1).

\bibliographystyle{abbrv}
\bibliography{references}

\begin{thebibliography}{10}

\bibitem{Ashbrook2003_gpsprediction}
D.~Ashbrook and T.~Starner.
\newblock {Using GPS to Learn Significant Locations and Predict Movement Across
  Multiple Users}.
\newblock {\em Journal of Personal and Ubiquitous Computing}, 7(5):275--286,
  October 2003.

\bibitem{Canzian2015MoodTraces}
L.~Canzian and M.~Musolesi.
\newblock {Trajectories of Depression: Unobtrusive Monitoring of Depressive
  States by means of Smartphone Mobility Traces Analysis}.
\newblock In {\em UbiComp'15}, Osaka, Japan, September 2015.

\bibitem{Chon2013}
Y.~Chon, E.~Talipov, H.~Shin, and H.~Cha.
\newblock {SmartDC: Mobility Prediction-based Adaptive Duty Cycling for
  Everyday Location Monitoring}.
\newblock {\em IEEE Transactions on Mobile Computing}, 13:1, 2013.

\bibitem{Consolvo2006Houston}
S.~Consolvo, K.~Everitt, I.~Smith, and J.~A. Landay.
\newblock Design requirements for technologies that encourage physical
  activity.
\newblock In {\em CHI'06}, Quebec, Canada, April 2006.

\bibitem{Consolvo2008}
S.~Consolvo, D.~W. Mcdonald, T.~Toscos, M.~Y. Chen, J.~Froehlich, B.~Harrison,
  P.~Klasnja, A.~Lamarca, L.~Legr, R.~Libby, I.~Smith, and J.~A. Landay.
\newblock {Activity Sensing in the Wild: a Field Trial of UbiFit Garden}.
\newblock In {\em CHI'08}, Florence, Italy, April 2008.

\bibitem{Fraga2013Emotiv}
T.~Fraga, M.~Pichiliani, and D.~Louro.
\newblock Experimental art with brain controlled interface.
\newblock In {\em Universal Access in Human-Computer Interaction. Design
  Methods, Tools, and Interaction Techniques for eInclusion}, pages 642--651.
  Springer, 2013.

\bibitem{Hafezi2015Ingestible}
H.~Hafezi, T.~L. Robertson, G.~D. Moon, K.~AuYeung, M.~J. Zdeblick, and G.~M.
  Savage.
\newblock {An Ingestible Sensor for Measuring Medication Adherence}.
\newblock {\em IEEE Transactions on Biomedical Engineering}, 62(1):99--109,
  2015.

\bibitem{Healey2005}
J.~Healey and R.~W. Picard.
\newblock {Detecting stress during real-world driving tasks using physiological
  sensors}.
\newblock {\em IEEE Transactions on Intelligent Transportation Systems},
  6(2):156--166, 2005.

\bibitem{Klasnja2012}
P.~Klasnja and W.~Pratt.
\newblock {Healthcare in the Pocket: Mapping the Space of Mobile-Phone Health
  Interventions}.
\newblock {\em Journal of Biomedical Informatics}, 45(1):184--198, 2012.

\bibitem{Lane2011b}
N.~D. Lane, T.~Choudhury, A.~Campbell, M.~Mohammod, M.~Lin, X.~Yang, A.~Doryab,
  H.~Lu, S.~Ali, and E.~Berke.
\newblock {BeWell: A Smartphone Application to Monitor, Model and Promote
  Wellbeing}.
\newblock In {\em Pervasive Health'11}, Dublin, Ireland, May 2011.

\bibitem{Lane2010}
N.~D. Lane, E.~Miluzzo, H.~Lu, D.~Peebles, T.~Choudhury, and A.~T. Campbell.
\newblock {A Survey of Mobile Phone Sensing}.
\newblock {\em IEEE Communications Magazine}, 48(9):140--150, 2010.

\bibitem{Lathia2013}
N.~Lathia, V.~Pejovic, K.~Rachuri, C.~Mascolo, M.~Musolesi, and P.~J. Rentfrow.
\newblock {Smartphones for Large-Scale Behaviour Change Intervention}.
\newblock {\em IEEE Pervasive Computing}, 12(3), 2013.

\bibitem{Lester2006Practical}
J.~Lester, T.~Choudhury, and G.~Borriello.
\newblock A practical approach to recognizing physical activities.
\newblock In {\em Pervasive Computing}, pages 1--16. Springer, 2006.

\bibitem{Lu2012}
H.~Lu, G.~T.~C. Mashfiqui~Rabbi, D.~Frauendorfer, M.~S. Mast, A.~T. Campbell,
  D.~Gatica-Perez, and T.~Choudhury.
\newblock {StressSense: Detecting Stress in Unconstrained Acoustic Environments
  using Smartphones}.
\newblock In {\em UbiComp'12}, Pittsburgh, PA, USA, September 2012.

\bibitem{Madan2010}
A.~Madan, M.~Cebrian, D.~Lazer, and A.~Pentland.
\newblock {Social Sensing for Epidemiological Behavior Change}.
\newblock In {\em UbiComp'10}, Copenhagen, Denmark, September 2010.

\bibitem{Mancinelli2000}
L.~Mancinelli, M.~Cronin, and W.~Sadée.
\newblock {Pharmacogenomics: the promise of personalized medicine}.
\newblock {\em AAPS Pharmsci}, 2(1):29--41, 2000.

\bibitem{Mehrotra2015NotifyMe}
A.~Mehrotra, M.~Musolesi, R.~Hendley, and V.~Pejovic.
\newblock {Designing Content-driven Intelligent Notification Mechanisms for
  Mobile Applications}.
\newblock In {\em UbiComp'15}, Osaka, Japan, September 2015.

\bibitem{Nasseh2012Adherence}
K.~Nasseh, S.~G. Frazee, J.~Visaria, A.~Vlahiotis, and Y.~Tian.
\newblock Cost of medication nonadherence associated with diabetes,
  hypertension, and dyslipidemia.
\newblock {\em American Journal of Pharmacy Benefits}, 4(2):e41--e47, 2012.

\bibitem{Osterberg2005Adherence}
L.~Osterberg and T.~Blaschke.
\newblock Adherence to medication.
\newblock {\em New England Journal of Medicine}, 353(5):487--497, 2005.

\bibitem{Pejovic2014b}
V.~Pejovic and M.~Musolesi.
\newblock {Anticipatory Mobile Computing for Behaviour Change Interventions}.
\newblock In {\em Workshop on Mobile Systems for Computational Social Science
  (MCSS'14)}, Seattle, WA, USA, September 2014.

\bibitem{Pejovic2014}
V.~Pejovic and M.~Musolesi.
\newblock {InterruptMe: Designing Intelligent Prompting Mechanisms for
  Pervasive Applications}.
\newblock In {\em UbiComp'14}, Seattle, WA, USA, September 2014.

\bibitem{Pejovic2015}
V.~Pejovic and M.~Musolesi.
\newblock {Anticipatory Mobile Computing: A Survey of the State of the Art and
  Research Challenges}.
\newblock {\em ACM Computing Surveys (CSUR)}, 47(3), 2015.

\bibitem{Perry2015CheckEngine}
T.~S. Perry.
\newblock Giving your body a" check engine" light.
\newblock {\em Spectrum, IEEE}, 52(6):34--84, 2015.

\bibitem{PersonalizedMedicine2014}
{Personalised Medicine Coalition}.
\newblock {The Case for Personalized Medicine}, 2014.

\bibitem{Rachuri2010}
K.~K. Rachuri, M.~Musolesi, C.~Mascolo, P.~J. Rentfrow, C.~Longworth, and
  A.~Aucinas.
\newblock {EmotionSense: A Mobile Phones based Adaptive Platform for
  Experimental Social Psychology Research}.
\newblock In {\em UbiComp'10}, Copenhagen, Denmark, September 2010.

\bibitem{Rasmussen2007Adherence}
J.~N. Rasmussen, A.~Chong, and D.~A. Alter.
\newblock Relationship between adherence to evidence-based pharmacotherapy and
  long-term mortality after acute myocardial infarction.
\newblock {\em Jama}, 297(2):177--186, 2007.

\bibitem{Rosen1985}
R.~Rosen.
\newblock {\em {Anticipatory Systems}}.
\newblock Pergamon Press, Oxford, UK, 1985.

\bibitem{Russell2003}
D.~M. Russell, P.~P. Maglio, R.~Dordick, and C.~Neti.
\newblock {Dealing with Ghosts: Managing the User Experience of Autonomic
  Computing}.
\newblock {\em IBM Systems Journal}, 42:177--188, 2003.

\bibitem{Scellato2011}
S.~Scellato, M.~Musolesi, C.~Mascolo, V.~Latora, and A.~T. Campbell.
\newblock {Nextplace: A spatio-temporal prediction framework for pervasive
  systems}.
\newblock In {\em Pervasive'11}, San Francisco, CA, USA, June 2011.

\bibitem{Sutton1998}
R.~S. Sutton and A.~G. Barto.
\newblock {\em {Reinforcement Learning: An Introduction}}.
\newblock MIT Press, Cambridge, MA, USA, 1998.

\bibitem{Tavris1997}
C.~Tavris and C.~Wade.
\newblock {\em {Psychology in Perspective}}.
\newblock Longman, New York, NY, USA, 1997.

\bibitem{Weiser1991}
M.~Weiser.
\newblock {The Computer for the 21st Century}.
\newblock {\em Scientific American}, 265(3):94--104, 1991.

\end{thebibliography}
\end{document}